\documentstyle[12pt]{article}
\begin{document}
\title{Magnetism of Neutron Stars and Planets}
\author{B.G. Sidharth$^*$\\
Centre for Applicable Mathematics \& Computer Sciences\\
B.M. Birla Science Centre, Adarsh Nagar, Hyderabad - 500 063 (India)}
\date{}
\maketitle
\footnotetext{$^*$Email:birlasc@hd1.vsnl.net.in; birlard@ap.nic.in}
\begin{abstract}
It is shown in this paper that recent results that below the Fermi temperature,
Fermions obey anomalous semionic statistics, could explain such apprently
diverse phenomena as the magnetism of Pulsars and White Dwarfs on the one hand
and earth like planets, on the other.
\end{abstract}
\section{Introduction}
It is known that Neutron stars or Pulsars have strong magnetic fields of
$\sim 10^8$ Tesla in their vicinity, while certain White Dwarfs have
magnetic fields $\sim 10^2$ Tesla. If we were to use conventional arguments
that when a sun type star with a magnetic field $\sim 10^{-4}$ Tesla contracts,
there is conservation of magnetic flux, then we are lead to magnetic fields
for Pulsars and White Dwarfs which are a few orders of magnitude less than
the required values\cite{r1}.\\
We will now argue, that in the light of recent results that below the Fermi
temperature, the degenerate electron gas obeys a semionic statistics, that is
a statistics in between the Fermi-Dirac and Bose-Einstein, it is possible to
deduce the correct magnetic fields for Neutron stars and White Dwarfs. Moreover
this will also enable us to deduce the correct magnetic field of a planet
like the earth.
\section{Anomalous Behaviour Below the Fermi Temperature}
In recent years, it has been realized that under specific conditions, for
example low dimensionality or sub Fermi temperatures, Fermions exhibit an
anomalous character - they obey statistics inbetween the Fermi-Dirac and
Bose-Einstein statistics\cite{r2,r3}.\\
Let us specifically consider the case of sub Fermi temperatures (cf.ref.\cite{r3}).
To notice the anomalous behaviour in a simple way we observe that in this case
as is known\cite{r4} the assembly fills up each and every single particle
energy level below the Fermi energy, with the Fermionic occupation number $1$.
The density of states in momentum space is given by $d^3p$, exactly as in the
case of Bosons. Whence we obtain the well known result
\begin{equation}
\epsilon_F = \frac{\hbar^2}{2m} \left(\frac{6\pi^2}{v}\right)^{2/3}\label{e1}
\end{equation}
where $\epsilon_F$ is the Fermi energy. The result for Phonons which obey
Bose-Einstein statistics is identical to equation (\ref{e1}) (cf.ref.\cite{r3}).\\
The anomalous behaviour can also be seen as follows: We have for the energy
density $e,$ in this case
\begin{equation}
e \propto \int^{p_F}_o \frac{p^2}{2m} d^3p \propto T^{2.5}_F\label{e2}
\end{equation}
where $p_F$ is the Fermi momentum and $T_F$ is the Fermi temperature. On the
other hand, it is known that \cite{r5,r6} in $n$ dimensions we have,
\begin{equation}
e \propto T^{n+1}_F\label{e3}
\end{equation}
(For the case $n = 3,$ (\ref{e3}) is identical to the Stefan-Boltzmann law).
Comparison of (\ref{e3}) and (\ref{e2}) shows that the assembly behaves with
the fractal dimensionality $1.5$.\\
Let us now consider an assembly of $N$ electrons. As is known, if $N_+$ is the
average number of particles with spin up, the magnetisation per unit volume
is given by
\begin{equation}
M = \frac{\mu (2 N_+ - N)}{V}\label{e4}
\end{equation}
where $\mu$ is the electron magnetic moment. At low temperatures, in the usual
theory, $N_+ \approx \frac{N}{2}$, so that the magnetisation given in
(\ref{e4}) is very small. On the other hand, for Bose-Einstein statistics we
would have, $N_+ \approx N$. With the above semionic statistics we have,
\begin{equation}
N_+ = \beta N, \frac{1}{2} < \beta < 1,\label{e5}
\end{equation}
If $N$ is very large, this makes an enormous difference in (\ref{e4}).
Let us use (\ref{e4}) and (\ref{e5}) for the case of Neutron stars.
\section{Magnetism of Neutron Stars and White Dwarfs}
In this case, as is well known, we have an assembly of degenerate electrons
at temperatures $\sim 10^7K$, (cf.for example \cite{r4}). So the considerations
of Section 2 apply. In the case of a Neutron star we know that the number
density of the degenerate electrons, $n \sim 10^{31}$ per c.c.\cite{r7,r8}. So
using (\ref{e4}) and (\ref{e5}) and remembering
that $\mu \approx 10^{-20}G,$ the magnetic field near the Pulsar is $\sim
10^{11}G \leq 10^8$ Tesla, as required.\\
As mentioned earlier some White Dwarfs also have magnetic fields. If the White
Dwarf has an interior of the dimensions of a Neutron star, with a similar
magnetic field, then remembering that the radius of a White Dwarf is about
$10^3$ times that of a Neutron star, its magnetic field would be $10^{-6}$
times that of the neutron star, which is known to be the case.\\
\section{Discussion}
It is quite remarkable that the above mechanism can also explain the magnetism
of the earth. As is known the earth has a solid core of radius of about
1200 kilometers and temperature about 6000 K\cite{r9}. This core is made up
almost entirely of Iron $(90 \%)$ and Nickel $(10 \%)$. It can easily be
calculated that the number of particles $N \sim 10^{48}$, and that the
Fermi temperature $\sim 10^5$. In this case we can easily verify using
(\ref{e4}) and (\ref{e5}) that the magnetic field near the earth's surface
$\sim 1G$, which is indeed the case.\\
It may be mentioned that the anomalous Bosonic behaviour given in (\ref{e5}) would imply a
sensitivity to external magnetic influences which could lead to effects
like magnetic flips or reversals.\\
To see this, we observe that the number of electrons, with spin aligned along
a magnetic field $B$ which is introduced, where,
$$B < < \epsilon_F/2\mu,$$
is given by (cf.ref.\cite{r4}), using Fermi-Dirac statistics,
$$N_+ \approx \frac{N}{2} (1+\frac{3\mu B}{2\epsilon_F})$$
That is, $\beta$ in (\ref{e5}) is given by
$$\beta \approx \frac{1}{2},$$
and the introduction of the field $B$, does not lead to a significant magnetic
field in (\ref{e4}). But, if as in Section 2, $\beta \ne \frac{1}{2}$, but
rather, $\beta > \frac{1}{2}$, then in view of the fact that $N$ is very
large, the contribution from (\ref{e4}) could be significant.\\
Indeed in the case of the earth
magnetic reversals do take place from time to time and are as yet not
satisfactorily explained\cite{r10}.

\end{document}